\documentclass[sigconf,dvipsnames]{acmart}

\AtBeginDocument{%
  \providecommand\BibTeX{{%
    \normalfont B\kern-0.5em{\scshape i\kern-0.25em b}\kern-0.8em\TeX}}}
    
\usepackage[utf8]{inputenc}
\usepackage{url}
\usepackage{colortbl}
\usepackage{caption}
\usepackage{balance}
\usepackage{xspace}
\usepackage[caption=false]{subfig}
\usepackage{graphicx}
\usepackage{verbatim}
\usepackage{bold-extra}
\usepackage{amsmath}
\usepackage{multirow}
\usepackage{listings}
\usepackage{boxedminipage}
\usepackage{soul}
\usepackage{enumitem}
\usepackage[normalem]{ulem}
\setlist{nolistsep,leftmargin=.5cm}
\usepackage{booktabs}
\usepackage{tabularx}
\usepackage{svg}
\usepackage{diagbox}
\usepackage{calc}
\usepackage{float}
\usepackage{multirow}
\usepackage[normalem]{ulem}
\useunder{\uline}{\ul}{}
\usepackage{amsmath}
\usepackage[most]{tcolorbox}
\usepackage{caption}
\usepackage{microtype} 
\usepackage{cleveref}

\makeatletter

\definecolor{mycolor}{RGB}{204,204,204}

\newboolean{showcomments}

\setboolean{showcomments}{true}

\ifthenelse{\boolean{showcomments}}
  {\newcommand{\nb}[2]{
    \fbox{\bfseries\sffamily\scriptsize#1}
    {\sf\small$\blacktriangleright$\textit{#2}$\blacktriangleleft$}
   }
   
  }
  {\newcommand{\nb}[2]{}
   
  }

\copyrightyear{2024}
\acmYear{2024}
\setcopyright{rightsretained}
\acmConference[ICSE 2024]{46th International Conference on Software Engineering}{April 2024}{Lisbon, Portugal}
\acmBooktitle{2024 IEEE/ACM 46th International Conference on Software Engineering (ICSE '24), April 14--20, 2024, Lisbon, Portugal}
\acmDOI{10.1145/3597503.3608139} 
\acmISBN{979-8-4007-0217-4/24/04}

\usepackage{xspace}

\begin{document} 

\title{
	Toward Rapid Bug Resolution for Android Apps
}

\author{Junayed Mahmud}
\email{junayed.mahmud@ucf.edu}
\affiliation{%
	\institution{University of Central Florida}
	\country{Orlando, FL, USA}
}

\renewcommand{\shortauthors}{J. Mahmud}

\begin{abstract}

Bug reports document unexpected behaviors in software, enabling developers to understand, validate, and fix bugs. Unfortunately, a significant portion of bug reports is of low quality, which poses challenges for developers in terms of addressing these issues. Prior research has delved into the information needed for documenting high-quality bug reports and expediting bug report management. Furthermore, researchers have explored the challenges associated with bug report management and proposed various automated techniques. Nevertheless, these techniques exhibit several limitations, including a lexical gap between developers and reporters, difficulties in bug reproduction, and identifying bug locations. Therefore, there is a pressing need for additional efforts to effectively manage bug reports and enhance the quality of both desktop and mobile applications. In this paper, we describe the existing limitations of bug reports and identify potential strategies for addressing them. Our vision encompasses a future where the alleviation of these limitations and successful execution of our proposed new research directions can benefit both reporters and developers, ultimately making the entire software maintenance faster.

\end{abstract}
\vspace{-2em}

\vspace{-3em}

\maketitle

\vspace{-1em}
\section{Introduction}
The popularity of desktop and mobile applications has ushered in a period of rapid growth in the software development landscape. To ensure that software reaches a high standard of quality, it necessiates substantial collaboration among developers, testers, and end-users. The software maintenance process is very complex, which often leads to the emergence of numerous bugs in practice. Software failures have, in fact, resulted in substantial financial losses, with billions of dollars lost by project owners because of these failures \cite{Charette2005}. Consequently, it becomes imperative to establish a robust strategy for \textit{identifying} software bugs, \textit{prioritizing} critical bugs, and \textit{resolving} them. Addressing critical bugs before software releases is paramount to avoiding financial losses for software companies and mitigating potential damages. One widely accepted approach to managing software failures is the creation of bug reports.

Bug reporting serves as a process for meticulously documenting and describing any unexpected behavior seen in software. Both \textit{non-contributors (e.g., end-users)} and \textit{contributors (e.g., developers)} generate bug reports when encountering any issues. Researchers have conducted qualitative studies, gathering insights from developers and reporters to delineate components that constitute high-quality bug reports \cite{Bettenburg2008GoodBR, Zimmermann2010}. These studies have elucidated how reporters usually describe \textit{observed behavior (OB)}, i.e., unexpected behaviors in software, \textit{expected behavior (EB)}, i.e., behaviors they expect, and \textit{steps to reproduce (S2R)} the bug. In addition, reporters add necessary screenshots, videos, stack traces, app versions, and system versions to provide comprehensive information. The most widely used bug-reporting platforms include Github Issue Tracker \cite{github-it}, Bugzilla \cite{bugzilla}, and Jira \cite{jira}. Unfortunately, a substantial number of existing bug reports lack the essential bug report components \cite{GitHub2016}.

The inadequacies of existing bug reporting systems are evident in their failure to bridge the \textit{lexical gap} between end-users and developers \cite{Moran2015, Zimmermann2010}. Existing bug reporting systems (i) offer \textit{limited guidance} on the contents and the presentation of those contents in bug reports and (ii) \textit{lack mechanisms} to provide feedback to reporters regarding the correctness and completeness of the bug report information \cite{song2022burt, song2022toward}. Consequently, due to the static nature of these bug-reporting interfaces, the responsibility of furnishing high-quality information falls on the shoulders of the reporters. One potential solution to mitigate this lexical gap is the adoption of interactive bug reporting. Building upon earlier research on the interactive question-answering system \cite{Ko2008}, conversational agents can play a pivotal role in enhancing the bug-reporting experience. 

The process of bug report management has been identified as a time-consuming endeavor \cite{murphy2013design, weiss2007long}. The difficulty is exacerbated when dealing with low-quality bug reports, as they often lack essential bug report components \cite{Bettenburg2008GoodBR, Chaparro2019, Chaparro2017, song2020bee}. Researchers have worked on automating text-retrieval (TR)-based bug localization, wherein an approach automatically \textit{retrieves and ranks buggy files} based on the likelihood of containing bugs \cite{davies2012using}. Prior research on automated bug localization concentrated textual content from bug reports, applying query reformulation strategies utilizing various information sources such as source code \cite{Rahman2018}. Moreover, researchers utilized execution information such as execution or stack traces \cite{Wen2016, youm2017improved}, code dependencies performing source code analysis \cite{dit2013feature}, or prior version information history from source code repositories \cite{youm2017improved} for improved TR-based bug localization. Distinctively, our research explores a novel resource in TR-based bug localization—graphical user interfaces (GUIs). GUIs offer latent patterns within their multiple versions of visual representations, such as pixel-based (\texttt{screenshots}) and metadata-based (\texttt{html, uiautomator}) based representations \cite{Moran:SANER22}. Leveraging high-level execution information, where code elements are directly linked with the UI, and low-level information, such as execution trace, we aim to enhance the ranking of the retrieved files from the source code repositories. Additionally, this GUI information contains textual content that can be integrated with bug reports to further boost the ranking of the buggy files.  

While interactive bug reporting and automated bug localization have made improvements, there remains a need to evaluate and enhance the quality of existing bug reports automatically. The advent of large language models (LLMs), such as ChatGPT \cite{chatgpt} on GPT-4, LLaMA \cite{llama}, PaLM \cite{chowdhery2022} has opened up opportunities to assess and improve the quality of the bug reports automatically. These LLMs can be utilized in identifying and extracting bug report elements from textual contents, and mapping them to app screens and GUI components, ultimately leading to the generation of high-quality bug reports and improving low-quality bug reports. 

Our research endeavors to enhance software maintenance processes by improving bug-reporting mechanisms and expediting bug resolution. We aim to pioneer the next-generation bug reporting mechanism by bridging the knowledge gap between reporters and developers, assisting developers in identifying bug locations, and enhancing low-quality bug reports by automatically incorporating essential bug report elements. Firstly, we designed the first interactive bug report system, a departure from conventional bug reporting systems such as Github Issue Tracker \cite{github-it} and Jira \cite{jira}. These existing systems offer limited guidance to bug reporters, lack mechanisms for verifying the quality of reported information, and provide no feedback on reporting bugs. In contrast, our interactive bug reporting system addresses these shortcomings by providing comprehensive support, assessing the quality of the written descriptions, and providing feedback to bug reporters. Secondly, identifying buggy source code locations presents a significant challenge for developers and researchers have proposed many automated techniques. Our contribution lies in being the first to apply automated bug localization techniques specifically to Android bug reports, leveraging GUI information to improve existing bug localization. Thirdly, we recognize the value of improving the quality of existing bug reports. This process involves the identification of essential components within bug reports, extracting these components, and associating them with specific GUI screens and components. In contrast to previous research efforts, we are the first to employ LLMs to accomplish this task.

\vspace{-1em}
\section{Current Research}
To enhance software maintenance tasks, the acquisition of high-quality bug reports and the automated identification of buggy code locations can significantly expedite bug understanding, reproduction, and resolution processes. 
The design of our research draws inspiration from successful results from the development of an interactive bug reporting system \cite{song2022toward,song2022burt}, as well as leveraging GUIs for automatically identifying buggy locations in source code  \cite{Mahmud2023}, improving the low-quality of bug reports using LLMs such as ChatGPT \cite{chatgpt}.

\subsection{An Interactive Bug Reporting System}
Our research instantiated with designing an interactive BUg ReporTing (BURT) system \cite{song2022toward, song2022burt}, which guides users while reporting bugs, assesses the quality of the bug report elements, provides feedback to the reporters, suggests graphical contents during reporting, and generates high-quality bug reports. The entire bug-reporting system comprises three components:

\subsubsection{\textbf{Natural Language Parser (NL)}}
BURT begins by labeling sentences from user-written descriptions and parses the sentences using a dependency parsing strategy. Each sentence is parsed according to a fixed format:
\begin{center}
	\texttt{[subject][action][object][preposition][object2]}
\end{center}
where the \textit{subject} can be a user or an app component, the \textit{action} can be actions like click, type, long-click, swipe, or pinch, and the \textit{object} denotes the entity on which the action is performed, with \textit{object2} serving as a secondary entity connected to the first by a \textit{preposition}.

\subsubsection{\textbf{Dialogue Manager (DM)}}
Our interactive bug reporting system initiates a dialogue with the user to gather essential bug report components such as OB, EB, and S2Rs. BURT suggests relevant app screens based on the information provided by the user, allowing users to select one or multiple screens. The system also asks if the user wants additional screen suggestions. The user can choose from additional screen suggestions or deny in which case, BURT prompts the user for an S2R description.

\subsubsection{\textbf{Report Processing Engine (RP)}}
BURT creates an app execution model that stores app interactions performed in the apps, and the app screens information in response to those interactions. This model represents app screens with GUI information as nodes and app interactions as edges. BURT performs quality verification by matching bug report descriptions with app screens and GUI components. From a collection of already mapped S2Rs to the app execution model, BURT suggests the next S2Rs.

\subsection{Leveraging GUIs for Bug Localization}
Our research also investigates the utilization of GUI interaction data, such as exercised GUI components during app traversal, GUI components within app screens, activity and window names, to enhance bug localization. The entire process is detailed in \cite{Mahmud2023}; however, we outline critical aspects below.

\subsubsection{\textbf{Mapping GUI Terms to GUI-Related Files}}
We explore how terms extracted from app screens and GUI components can be employed in bug localization. The GUI terms are appended to the query reformulation and also utilized to re-rank GUI-related files. We retrieve GUI-related files through three strategies: (i) matching GUI activity and window names extracted from \texttt{ViewServer} with corresponding .java class names, (ii) identifying files containing event listeners related to specific GUI components using their resource ids, and (iii) retrieving GUI-related files based on the presence of information about exercised GUI components in different app screens during bug reproduction.

\subsubsection{\textbf{Text-Retrieval Augmentation Methods}}
In our study, we employed two specific techniques in addition to baseline techniques for retrieving relevant source code files relevant to bug reports.

\textbf{Reformulating Queries using GUI terms.} 
Our initial text retrieval-based augmentation approach involves the reformulation of queries through two distinct strategies: (i) \textbf{Query Expansion:} This method involves appending pertinent GUI-related information to the bug report. Essentially, we enrich the query with additional context from the GUIs. (ii) \textbf{Query Replacement:} In this technique, we entirely replace the content of bug reports with relevant GUI-related information, which is then used as the new query.

\textbf{Re-Ranking using GUI-Related Files.} 
Our next augmentation approach is to explore various re-ranking strategies based on GUI information. These strategies encompass: (i) \textbf{Filtering:} We filter out all files that are not directly related to the GUI, ensuring that only GUI-relevant files remain in the ranked list. (ii) \textbf{Boosting:} In this strategy, files matching the GUI information are given priority and placed at the top of the ranked list.
	(iii) \textbf{Filtering + Boosting:} This final strategy combines both filtering and boosting. Boosted files constitute a subset of the filtered files, and they are positioned at the top among all the filtered files.

\subsection{Improving Bug Reports Quality using LLMs}
Given the success of large language models like ChatGPT \cite{chatgpt}, an alternative direction is to enhance the quality of existing bug reports by leveraging this LLM. This strategy involves performing the following steps:

\subsubsection{\textbf{Identification of Bug Report Components}}
Our initial objective here is to identify the components of bug reports, including OB, EB, and S2R, using various prompt engineering strategies \cite{promptEngineering}. Initially, we create a set of ground truth labels by identifying OB, EB, and S2R employing an open-coding strategy \cite{Miles2013} and prompt ChatGPT to identify necessary components. Remarkably, even a single sentence prompt (\textit{e.g., You will identify the sentences that describe the Observed Behavior of the app (OB), the Expected Behavior of the app (EB), and the Steps to Reproduce the bug (S2Rs) of the bug report.}) can successfully elicit identification without the need for specific contexts. We evaluate the generated responses quantitatively using standard metrics such as precision, recall, F1 score, and accuracy and qualitatively investigate the generated responses to understand where ChatGPT fails. Based on our findings, we identify cases where prompts fail to identify components and update prompts addressing issues found in the current prompt.  

\subsubsection{\textbf{Execution Model and S2R Resolution}}
In this phase, we generate an execution model of the application using established research strategies \cite{Chaparro2019, song2022burt, song2022toward}. We also extract actions, primitives, and objects \cite{Feng2023} using prompt engineering. These objects are mapped to relevant GUI screens and components using the app execution model.

\subsubsection{\textbf{Quality Report Generation}}
The most crucial component in bug reports is the S2Rs. Our approach rigorously examines each S2R within the bug report to determine if it should be segmented into multiple sentences, following a practice established in AdbGPT \cite{Feng2023}. Furthermore, after mapping sentences to specific app screens and GUI components and executing steps, we identify any missing app screens and interactions that occurred during bug reproduction. This meticulous process allows us to validate and generate high-quality bug reports.

\vspace{-1em}
\section{Empirical Results}
\label{sec:results}

\subsection{BURT Performance}
In order to assess the performance of BURT, we conducted a study involving participant recruitment and analysis of responses. After filtering out unusable data, we focused on the input provided by 18 participants. Participants received training on how to utilize the tool and were shown recordings of bug reproductions. Our analysis involved comparing BURT with ITRAC, a conventional web form for bug reporting that shares similarities with platforms like Github Issue Tracker \cite{github-it} and Jira \cite{jira}. The key findings from our evaluation are as follows: (1) Reporters \textit{favored the suggestions} offered by BURT over ITRAC. However, they expressed a desire for more precise and well-phrased suggestions. (2) A majority of reporters found BURT to be \textit{user-friendly and easy to navigate.} (3) BURT's components were deemed \textit{sufficient}, but improvements in accuracy are still required. (4) Bug reports collected using BURT were of \textit{higher quality} compared to those gathered through ITRAC. Nevertheless, there is room for enhancement in the descriptions of both OB and EB.

\subsection{GUI Bug Localization Performance}
To evaluate how GUI interaction data can enhance bug localization, we employed the AndroR2 dataset \cite{Wendland2021, Johnson2022}, which consists of \textbf{180 bug reports} across various categories, including \textit{output-, cosmetic-, navigation-, and crash-}related bugs. After filtering out bug reports, such as those where the bug solely involves .xml files or Kotlin code, etc., we were left with 80 bug reports where at least one buggy .java file exists in the ground of the buggy locations.
In our study, we applied reformulation and re-ranking strategies in four baseline bug localization techniques: BugLocator \cite{Zhou2012}, SentenceBERT \cite{reimers2019}, UniXCoder \cite{guo2022unixcoder}, and Lucene \cite{apacheLucene} for \textbf{657 different configurations} for each baseline.
Our key findings are as follows: (1) Leveraging GUI information from the \textit{buggy screen and the two preceding screens} resulted in the highest bug localization performance. This suggests that not only the buggy screen but also the screens preceding it have an impact on identifying the source code files responsible for the bug.
(2) \textit{Filtering} proved beneficial in only a few cases, whereas \textit{boosting} was advantageous in most scenarios. \textit{Query replacement} was not effective for improving bug localization, but \textit{query expansion} consistently improved performance across all baselines except for SentenceBERT. The best results were obtained when both \textit{filtering and boosting} were employed.
(3) The most optimal configuration involved using \textit{filtering and boosting simultaneously} in terms of improvements in Hits@10 over baselines.

\subsection{ChatGPT-based Bug Reporting Performance} 
Our overarching objective was to assess the capabilities of ChatGPT in generating high-quality bug reports using existing prompt engineering strategies. We began by providing minimal information in the prompt, identifying instances where ChatGPT faltered, categorizing these issues, and adjusting prompts based on our findings.
Our current progress includes the identification of bug report components through prompt engineering. A few instances where ChatGPT struggled include: (i) treating app/device version information as the OB (e.g., App Version: 1.5.8.). (ii) mislabeling an S2R as an EB when explicit actions on a GUI component are mentioned (e.g., Press Save).
At present, ChatGPT has demonstrated promising performance in terms of precision, recall, and F1 score in identifying bug report elements.
Our upcoming steps involve completing the S2R resolution phase and generating high-quality bug reports. We anticipate completing these tasks within three months and subsequently publish our findings.

\vspace{-0.5em}
\section{Proposed Research}
\label{sec:proposed-research}
Our objective is to enhance the overall software maintenance cycle by enhancing the process of bug report management. We aim to achieve this by generating high-quality bug reports, automatically locating buggy source code locations utilizing various resources, improving existing bug reports quality and potentially automating program repairs. The ultimate goal is to expedite bug report maintenance and create software that is free of bugs. To accomplish this, we propose to tackle the following research challenges:

\subsection{Creating Best Quality Bug Reports}
In our research, we introduce BURT \cite{song2022burt, song2022toward}, an interactive bug system designed to produce high-quality bug reports. However, despite its capabilities, several challenges remain to be addressed. Firstly, BURT encounters difficulties when mapping user-written sentence descriptions to the corresponding application screens, particularly when the sentence quality is low. Secondly, the feedback provided by BURT does not consistently meet high-quality standards. 

To overcome these limitations, we propose leveraging a combination of LLM with BURT. ChatGPT demonstrates promising performance in understanding low-quality sentences. By fine-tuning ChatGPT using various types of Android bug reports, experimenting with different prompt engineering strategies, and incorporating an app execution model, we aim to enhance our ability to map sentences to app screens and GUI components. This will result in more comprehensible bug reports and the identification of any missing steps in the steps-to-reproduce (S2R) section.  

\subsection{Automated Method-Based Bug Localization}
Most previous research has focused on identifying source code files given bug reports. Our empirical studies exhibit the advantages of employing GUIs in bug localization. By leveraging GUI information obtained from \texttt{ViewServer} and identifying potential \texttt{event listeners}, we can also improve method-based bug localization. This approach offers a fine-grained level of granularity, which can be especially beneficial for developers.  

\subsection{Integration to Industrial Applications}
Our previous research has highlighted the need for continuous improvement in bug reporting systems and the automated identification of potential buggy locations. Rather than relying solely on traditional metrics commonly used in existing techniques, we plan to conduct studies in industrial settings. After designing an improved bug reporting system, we intend to engage end-users and developers from industries in reporting bugs and evaluate their experiences.
Furthermore, we also aim to integrate our automated bug localization approach into real-world industrial bug-reporting systems. This will enable us to evaluate whether it facilitates quicker and more accurate identification of bug locations, ultimately making developers' tasks more manageable. We will actively seek feedback from developers to refine and enhance our tools.

\subsection{Automated Program Repair using GUIs}
Our research demonstrates the potential of GUIs in automatically identifying buggy source code locations. Following the identification of buggy GUI components and an understanding of the expected app behavior as described in the bug report, we can formulate strategies for addressing GUI-related bugs in source code. This approach enables us to repair programs with minimal developer involvement, potentially resolving software bugs more efficiently.

\vspace{-0.5em}
\section{Expected Contributions}
\label{sec:expected-contributions}
Our research has identified the existing limitations within the bug reporting system. While various phases are involved in bug reporting, our primary focus centers on improving the quality of bug reports and locating software bugs with the objective of comprehending, addressing, and resolving bugs promptly. Regarding the current utilization of tools and practices for the automated identification of buggy code files, we foresee the need for significant improvements in the near future to enhance software quality and minimize bugs. Presently, conventional practices entail employing low-quality bug reporting systems and manually identifying and fixing bugs, a process that consumes a substantial amount of developers' time as they need to understand, reproduce, and fix these issues. Our research on interactive bug reporting systems and the potential utilization of multiple resources, such as GUIs, in automated bug localization, leveraging LLM in improving bug report quality introduces a novel direction in this research. By integrating both textual and visual components, we have the opportunity to revolutionize thinking and apply these innovations to industrial applications. Our incorporation of these emerging techniques holds the promise of introducing a new dimension to the bug reporting system, one that significantly mitigates the knowledge gap between bug reporters and developers. We envision a future where addressing bugs will require minimal involvement from software developers, thus expediting bug report management and creating high-quality software.

\vspace{-1em}
\section{Conclusion}
We introduce current practices and identify limitations in the realm of bug reporting. Therefore, we have envisioned a roadmap for future endeavors aimed at generating high-quality bug reports and streamlining the responsibilities of both developers and reporters. Additionally, we have studied the utilization of rich resources such as GUIs and adapting advanced techniques such as LLMs to fix bugs quicker and improve the quality of existing bug reports. The success of our research could potentially usher in a new era of bug reporting, offering innovative solutions for the efficient maintenance of software systems.


\balance

\bibliographystyle{ACM-Reference-Format}
\bibliography{references}

\end{document}